\documentclass[journal=jacsat,manuscript=supporting-information]{achemso}
\usepackage{tocloft} 
\usepackage{setspace} 
\usepackage{achemso}
\setkeys{acs}{articletitle = true}
\usepackage[T1]{fontenc}       
\usepackage{graphicx}
\usepackage{dblfloatfix}
\usepackage{placeins}
\usepackage{physics,amsmath}
\usepackage{xcolor}
\usepackage{bm}
\usepackage{float}
\usepackage{makecell}
\usepackage{caption}
\usepackage{multirow}
\usepackage{array}
\usepackage{geometry}
\usepackage{booktabs, tabularx} 
\newcolumntype{C}{>{\centering\arraybackslash}X}

\newcommand*{\tlcj}[1]{\textcolor{black}{ #1}}
\newcommand*{\sg}[1]{\textcolor{black}{ #1}}

\author{Sayan Ghosh}
\affiliation{Solid State and Structural Chemistry Unit, Indian Institute of Science, Bangalore, Karnataka 560012, India}
\author{Amitav Sahu}
\affiliation{Present Address: Division of Chemical Physics and NanoLund, Lund University, Box 124, 221 00 Lund, Sweden}
\author{Stephanie Gonzalez-Migoni}
\affiliation{Zernike Institute of Advanced Materials, University of Groningen, Nijenborgh 3, 9747 AG Groningen, Netherlands.}
\author{Thomas L. C. Jansen}
\email{t.l.c.jansen@rug.nl}
\affiliation{Zernike Institute of Advanced Materials, University of Groningen, Nijenborgh 3, 9747 AG Groningen, Netherlands.}
\author{Vivek Tiwari}
\email{vivektiwari@iisc.ac.in}
\affiliation{Solid State and Structural Chemistry Unit, Indian Institute of Science, Bangalore, Karnataka 560012, India}

\title{Prominent Signatures of Energy Transfer in Action-Detected Spectra of a Cyanobacterial Photosynthetic Protein}

\begin{document}
\maketitle

\begin{abstract}
    Action-detected two-dimensional electronic spectroscopy (A-2DES) could potentially be a versatile chemical tool with applicability across a range of photophysical observables such as photocurrent, photoionization, or fluorescence. However, a prominent absence of excited state energy/charge transfer dynamics signals in archetypal photosynthetic proteins has suggested severe limitations of A-2DES in probing large aggregates where sensitivity to excited state dynamics is proposed to go down as $1/N$, where $N$ is the aggregate size. We report measurements of energy transfer dynamics in a cyanobacterial protein through both conventional and fluorescence 2DES (F-2DES), where the dynamics reported by F-2DES is quite prominent and comparable to that measured by conventional 2DES. Analysis of our experiments combined with coarse-grained simulations of the spectra suggest that the $1/N$ limit argument, which assumes infinitely fast intra-exciton manifold equilibration, is modified in case of cyanobacterial proteins because of slow annihilation. Our results suggest that action detection may in fact be well-suited to probe exciton diffusion across weakly coupled systems.
\end{abstract}

\begin{figure*}[h]
 \centering
 \includegraphics[width=14cm]{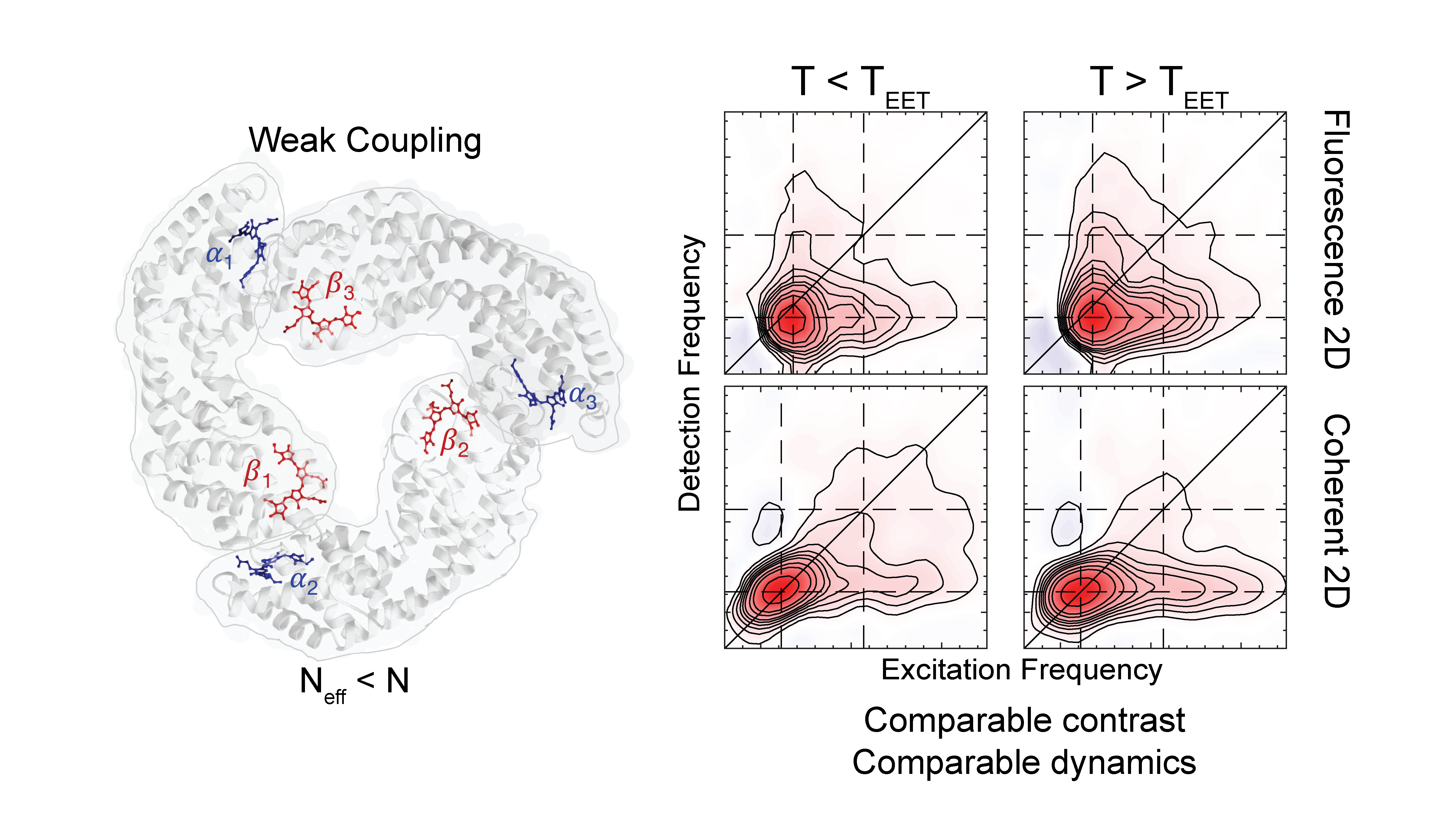}
 \label{TOC Graphic}
\end{figure*}


\section*{INTRODUCTION}
Two-dimensional electronic spectroscopy \cite{Jonas2003} (2DES) maps pump-probe waiting time ($T$) dynamics of a system as a 2D contour map along the excitation and detection frequency axes, thus providing unprecedented insights into ultrafast phenomena spanning chemistry and physics, such as photosynthetic energy and charge transfer \cite{JonasARPC2018}, inter-layer excitons in layered inorganic semiconductors \cite{Policht2023} and electronic relaxation pathways in hybrid light-matter states \cite{Zigmantas2021}. Among these classes of spectroscopies, action-detected 2DES \cite{Tiwari2021} has gained prominence in recent years as a highly sensitive \cite{Sahu2023} and versatile spectrometer with broad applicability. Such demonstrations include 2D photocurrent spectroscopy \cite{Karki2014,Nardin2013} of photovoltaic devices, spatially-resolved 2D fluorescence spectroscopy\cite{Tekavec2007} of photosynthetic cells \cite{Tiwari2018} and layered materials \cite{Li2021}, and 2D photoionization spectroscopy\cite{Bruder2018,Roeding2018} of dilute gases. 

Due to incoherent detection, it has long \cite{Lott2011,Marcus2012} been understood that mixing of pump and probe induced populations during detector time integration, for example, due to exciton-exciton annihilation (EEA), can lead to undesired signal pathways such as multiple excited state absorption (ESA) signal pathways with opposite sign. EEA and mutual cancellation between these ESA signals explain the strong positive $T = 0$ fs 2D cross-peaks (CPs) in covalently coupled synthetic dimers\cite{Lott2011,Maly2018,Kuhn2020,Tiwari2018b,Maly2020} as simply ground state bleach (GSB) signals that need not reflect the extent of coherent coupling between the chromophores. 

The situation in the case of large photosynthetic aggregates, such as the archetypal LH2 photosynthetic antenna protein\cite{Tiwari2018,Kunsel2019,Karki2019} from purple bacteria, is even more complex. Such systems are known for highly efficient EEA \cite{Pullerits2001}, one of the photoprotection mechanisms\cite{Blankenship2002}. Perfect EEA in photosynthetic antennas implies perfect mutual cancellation between the ESA signals, thereby leading to incoherent GSB signals between uncorrelated sites dominating the stimulated emission (SE) signal due to energy transfer between coupled sites. In an aggregate with $N$ perfectly coupled sites, it has been proposed\cite{Bolzonello2023} that the incoherent GSB signal dominates the coherent SE signals by $1/N$, that is, the “$1/N$ limit”. The $1/N$ limit explains\cite{Javed2024} the stunning lack of pump-probe waiting time dynamics in the LH2 protein. Thus, the efficacy of highly promising action-based 2DES approaches has recently come under significant scrutiny, with several alternative strategies being proposed, including gated fluorescence detection \cite{Maly2018,Kunsel2019}, two-dimensional fluorescence excitation \cite{Hauer2023}, and, more recently, approaches that exploit spectro-temporal symmetries to suppress incoherent background contributions \cite{Maly2025}.

It should be noted that the $1/N$ limit argument is inherently combinatorial in nature and assumes that exciton transfer within a manifold of $N$ coupled sites, and therefore EEA, is infinitely faster than radiative relaxation. Consequently, the cancellation between incoherent and coherent ESA signal pathways remains essentially perfect with changing pump-probe delay for sizable $N$. Both assumptions need not be valid. For instance, in their work on porphyrin dimers, Marcus and co-workers indeed discuss\cite{Lott2011,Marcus2012} that mutual cancellations between the ESA signals can depend on electronic relaxation during the pump-probe waiting time. In this work, we combine two-dimensional electronic spectroscopy (2DES) with both coherent and fluorescence-based (incoherent) detection, alongside coarse-grained simulations of spectral and dynamical behavior, to demonstrate that photosynthetic antennas exhibiting slower inter-site exciton transfer -- such as the cyanobacterial antenna protein studied here -- can display pronounced signatures of excited-state energy transfer in F-2DES measurements. Notably, these signatures closely resemble those observed in coherent 2DES (C-2DES). Contrary to the current challenges regarding the efficacy of F-2DES for photosynthetic systems with efficient EEA, our observations suggest action-detected spectroscopies may indeed be well suited to probe exciton diffusion across weakly coupled assemblies. Strongly coupled multi-chromophoric systems, such as the LH2 antenna protein, represent particularly challenging cases for observing energy transfer using F-2DES.

\section*{RESULTS}
Allophycocyanin (APC) is a highly-fluorescent\cite{Gantt1978} cyanobacterial antenna protein\cite{Blankenshipbook,Gantt1978} which forms the core of the phycobilisome antenna complexes of red algae and cyanobacteria, and funnels excitation toward the photosystem reaction centers. An $\alpha\beta$ APC monomer is comprised of $\alpha$ and $\beta$ polypeptide sub-units where the Cys84 residue of each sub-unit is covalently bound to a phycocyanobilin (PCB) chromophore. The APC monomer assembles as a trimeric protein complex, denoted as ($\alpha\beta$)$_3$ and shown in Figure S1. The $\alpha$ and $\beta$ PCBs within a given $\alpha\beta$ protein sub-unit are separated by $\sim$50 \r{A} and weakly coupled. However, the $\alpha$ and $\beta$ PCBs on the neighboring protein sub-units are separated by $\sim$21 \r{A}, energetically non-degenerate with weak excitonic coupling\cite{Womick2009,Womick2011}, and yet undergo rapid $\sim$220 fs downhill energy transfer\cite{Beck1998,Womick2009} which has been attributed\cite{Womick2009,Womick2011} to strong vibronic couplings. The neighboring $\alpha\beta$ PCB pairs are only weakly coupled and treated independently\cite{Womick2009,Womick2011} in the vibronic exciton models for energy transfer. APC thus provides an interesting contrast to the LH2 antenna protein of purple bacteria, where moderate to strong electronic couplings within the B800 and B850 rings\cite{ScholesFleming2000}, near unity yields of EEA\cite{Pullerits2001}, and rapid intra-ring electronic delocalization and downhill inter-ring energy transfer are well known. Recently F-2DES and C-2DES measurements of energy transfer in the LH2 protein revealed\cite{Javed2024} a striking lack of dynamics in the F-2D spectra. We contrast this by probing the excited-state dynamics of ($\alpha\beta$)$_3$ APC trimers using the F-2DES and C-2DES approaches. \\

Both spectrometers, fluorescence-detected\cite{Sahu2023} and coherent 2DES\cite{Bhat2023,Thomas2023} have been described in detail in our previous works and provide absorptive 2DES spectra as a function of the pump-probe waiting time $T$. Figures S2,S3 give a brief description of the setup and data collection details. In the C-2DES approach, a pump pulse pair induces absorptive changes in the sample that are recorded as a function of their mutual delay, $t_1$. These changes are probed at varying pump-probe delays $T$ for all probe detection wavelengths $\lambda_t$ or frequency $\omega_t$. A third-order oscillating macroscopic polarization is generated in the sample that starts coherently radiating with timing set by the interaction of the sample with the third (probe) pulse. The radiated electric field is homodyned with the probe, and the resulting 2D signal is recorded as $S(t_1 ,T, \lambda_t)$, where Fourier transform along $t_1$ provides the excitation frequency axis $\omega_1$ and 2D spectra $S(\omega_1 ,\omega_3; T)$. In contrast, in the F-2DES experiment, a fourth pulse, after a delay $t_3$ relative to the third pulse, can project the oscillating third-order polarization onto an excited state population which evolves orders of magnitude longer than the oscillating macroscopic polarization during a time interval $t_4$, leading to observables $O(t_1 ,T, t_3, t_4)$ such as photocurrent\cite{Karki2014,Nardin2013}, photoionized fragments\cite{Bruder2018,Roeding2018} or fluorescence\cite{Tekavec2007}. In our phase-modulation approach\cite{Tekavec2007} to white-light F-2DES\cite{Sahu2023}, the non-linear population arising due to the action of 4 pulses is separated by phase-modulation of each pulse at a unique radio-frequency $\Omega_i$ and detected using phase sensitive lock-in detection at two frequency filtered linear combinations $\mp \Omega_{12} + \Omega_{34} $, where $\Omega_{ij}$ is $\Omega_i - \Omega_j$, simultaneously by two independent channels of a lock-in amplifier (Zurich Instruments, HF2LI). Note that the experiment is blind to the evolution of the population during $t_4$ and time-integrates this interval. This implies that any incoherent mixing between pump and probe populations, denoted by $\langle{\psi_1}|{\psi_2}\rangle \bigotimes \langle{\psi_3}|{\psi_4}\rangle $ , where ${\psi_i}$ denotes first-order wave packets induced by the $i^{th}$ pulse and $\bigotimes$  denotes incoherent mixing of populations due to EEA for instance, will also appear at the frequency-filtered linear combinations.

\begin{figure*}[h]
 \centering
 \includegraphics[width=14cm]{APC_figure1_v6.png}
 \caption{(a) Absorptive F-2D spectra of the APC trimer. Each spectrum is normalized to its maximum value. The contours are drawn in 5\% intervals up to 30\%, 10\% intervals between 30\% and 50\%, and 25\% intervals between 50\% and 100\%. Black dashed lines are drawn at 16138, 15357 cm$^{-1}$ along $\omega_1$ and 15226 cm$^{-1}$ along $\omega_3$, at approximate locations of the upper and lower excitons. The top panel on the $T$ = 100 fs spectrum shows the Fourier transform fluorescence excitation (FT-FLEX) spectrum while the side panel shows the FT-FLEX and emission (EMS) spectra overlaid. (b) Integrated area of the horizontal band (shown in $T$ = 450 fs 2D) is plotted against excitation frequency for different $T$ points (from 100 fs (blue) to 450 fs (red) with 50 fs steps). (c) Area enclosed by the horizontal band and the vertical lines in the $CP_L$ region is integrated and plotted for different $T$. The integrated area is normalized to 1 for the initial $T$ point to directly reflect the $\sim44\%$  change. }
 \label{fig1}
\end{figure*}

\subsection*{Fluorescence-detected 2DES and pump-probe experiments}
Cross-linked APC (AnaSpec) was diluted in a 0.03 M phosphate buffer, maintaining a pH of 7.3. Cross-linking between the $\alpha\beta$ APC monomers is reported\cite{Glazer1987} to significantly enhance the trimer stability compared to its native counterpart while retaining the linear spectroscopic characteristics. Linear absorption (Figure S4\sg{d}) spectra showed minimal changes between before and after the experiments. The lower peak at 15357 cm$^{-1}$ (651 nm) is consistent with the lower exciton in APC that is known to arise in $(\alpha\beta)_3$ trimer due to Coulomb coupling between $\alpha$ and $\beta$ PCBs of adjacent $\alpha\beta$ sub-units.\\

Figure~\ref{fig1}a shows the absorptive F-2D spectra of APC at representative pump-probe delays until approximately two times the energy transfer timescale (\textit{vide infra}). The Fourier transform fluorescence excitation (FT-FLEX) spectrum is shown in the top panel of $T =$ 100 fs F-2D spectrum while the FT-FLEX overlaid with emission (EMS) is shown in the left panel. The upper diagonal peak ($DP_U$) is $\sim$ 3$\times$ weaker than the main diagonal peak ($DP_L$) and consistent with the expectation from the F-2D spectrum reconstructed from the linear FT-FLEX spectra (Figure S4\sg{a}). The $\sim13\%$ negative region near $DP_U$ arises due to the ringing baseline distortion associated with Fourier transform of the Heaviside function at the $t_{21} \ge 0$ and $t_{43} \ge 0$ boundary. Figure~\ref{fig1}b shows the growth in the lower cross peak ($CP_L$) region after integrating a horizontal band along the lower exciton marked in Figure~\ref{fig1}a ($T$ = 450 fs), and the area enclosed by the horizontal and vertical lines is plotted in Figure~\ref{fig1}c. The $CP_L$ region shows a prominent growth of $\sim44\%$. In comparison, the LH2 complex was reported\cite{Javed2024} to show a barely discernible growth of $\sim4\%$ in the $CP_L$ region and found to be approximately consistent with the predictions\cite{Bolzonello2023} of the $1/N$ limit. A theoretical comparison with the limiting case of infinitely fast EEA rate, compared to other electronic relaxation processes, is shown later in Figure~\ref{fig5}. It is important to note that, despite an approximately 44\% increase in $CP_L$, the F-2D spectrum of APC exhibits no pronounced contrast. This behavior is similar to the lack of amplitude contrast reported in the F-2D spectra of LH2\cite{Javed2024}. In LH2, strong contributions from the oppositely signed ESA signal generate a time-evolving amplitude contrast, as explained later in Figure~\ref{fig5}d, between the $CP_L$ and $DP_U$ regions in the C-2D spectra. For example, see Figure~1 of ref.\cite{Javed2024}. However, such ESA contributions are absent in the F-2D spectra of LH2, resulting in the observed suppression of amplitude contrast. The case of APC is investigated further in Figure~\ref{fig3} and Figure~\ref{fig5}.

\begin{figure}[h!]
\centering
  \includegraphics[width=7cm]{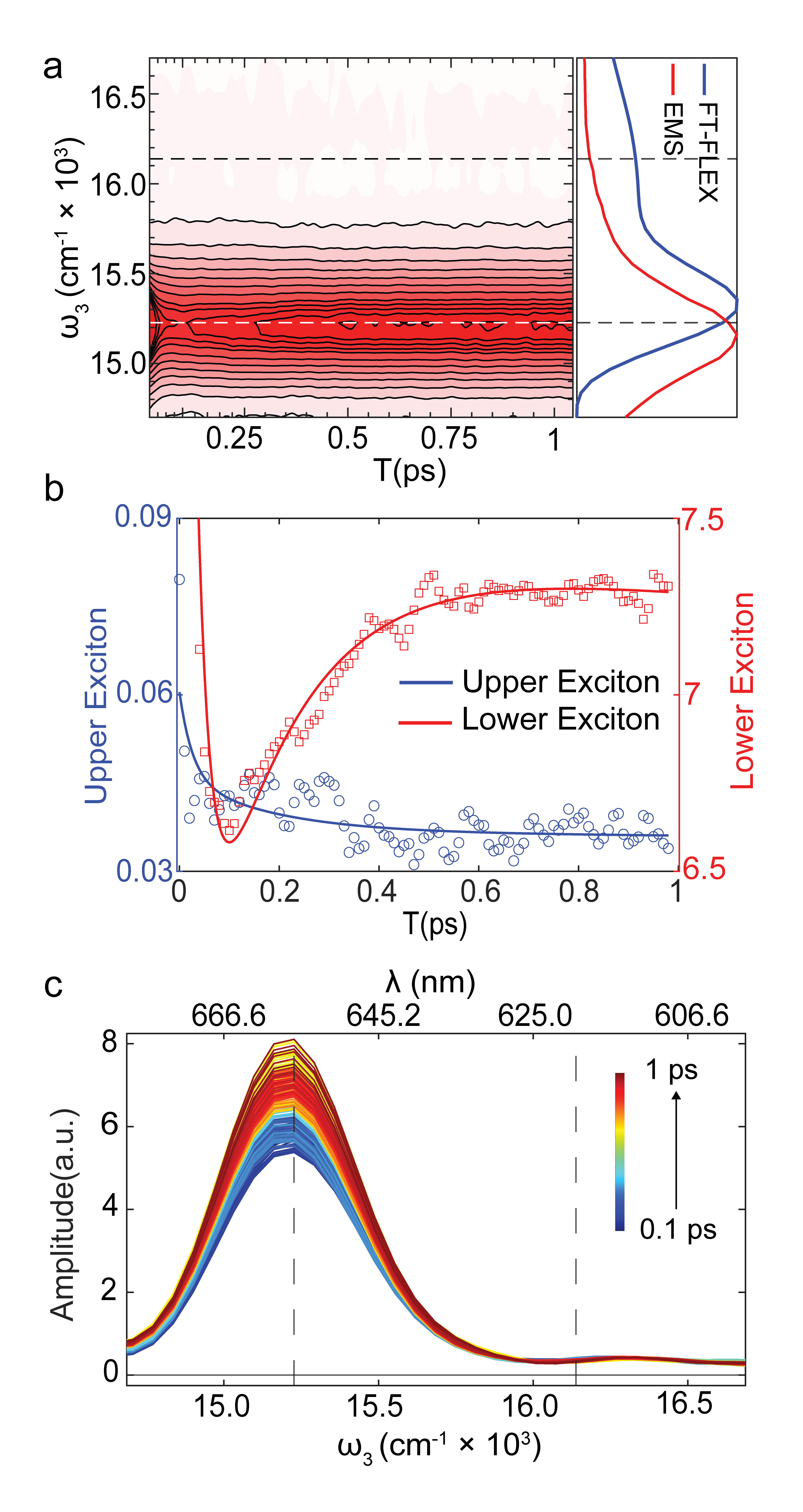}
  \caption{(a) Absorptive F-PP spectra of APC. Contours are drawn at 10\% intervals up to 80\% and 5\% intervals between 80\% to 100\%. Black dashed lines are drawn at 16138 and 15226 cm$^{-1}$ (620 and 656 nm, respectively). (b) The global fit of the F-PP data with a tri-exponential function gave the time constant of 182$\pm${1} fs. The data slices at the upper and lower exciton detection frequencies highlight the concomitant decay and rise of the upper and lower excitons, respectively. (c) F-PP spectra from panel (a) at different $T$ points, normalized at the upper exciton. Blue to red show $T$ range from 0.1 ps to 1 ps with 5 fs step, which reflect $\sim$50\% rise in the lower exciton. }
  \label{fig2}
\end{figure}

To capture the time constants for energy transfer with better signal-to-noise ratio, more $T$ points are needed for which a spectrally resolved pump-probe experiment is much faster because the excitation time delay $t_{21}$ is fixed at zero and only $T$ and detection time delay $t_{43}$ are scanned. The resulting fluorescence-detected pump-probe (F-PP) spectrum is shown in Figure~\ref{fig2}a where the peak locations are consistent with the F-2D spectra in Figure~\ref{fig1}. Note that the F-PP data is collected independently of the F-2D experiment and not derived from it. Figure~\ref{fig2}b shows a concomitant rise and decay at the location of the lower and upper excitons, respectively. The initial drop in the PP data is observed within the pulse envelope region and attributed to decay of non-time ordered signal pathways within the pulse overlap region \cite{Javed2024}. Global exponential fit of the data reveals a 182$\pm${1} fs component that is consistent with previous reports\cite{Womick2009,Womick2011} and assigned to electronic energy transfer (EET) between the upper and lower excitons formed by the $\alpha$ and $\beta$ PCB pigments of adjacent $\alpha\beta$ monomers of the $(\alpha\beta)_3$ trimer. Similar to F-2D $CP_L$ region, the growth of the lower exciton is prominently seen in the F-PP data. Figure~\ref{fig2}c follows the same methodology as Figures~\ref{fig1}b,c and plots the time slices of the F-PP data in Figure~\ref{fig2}a with normalization at the upper exciton, to show a $\sim$50\% amplitude rise in the lower exciton region.

\begin{figure*}[h!]
 \centering
 \includegraphics[width=14cm]{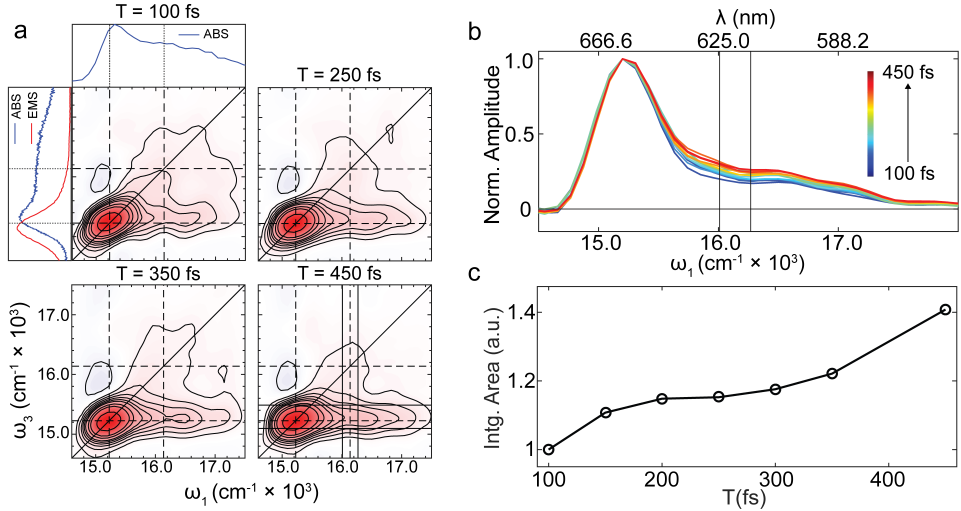}
 \caption{(a) Absorptive C-2D spectra of the APC trimer. Each spectrum is normalized to its maximum value, and the contours are drawn in 5\% intervals up to 30\%, 10\% intervals between 30\% and 50\%, and 25\% intervals between 50\% and 100\%. Black dashed lines are drawn at 16138 cm$^{-1}$ and 15357 cm$^{-1}$ along $\omega_1$ and 15226 cm$^{-1}$ along $\omega_3$, at approximate locations of the upper and lower excitons. The top panel on the $T$ = 100 fs spectrum shows the absorption spectrum (ABS) while the side panel shows the ABS and EMS spectra overlaid in blue and red color, respectively.(b) Integrated area of the horizontal band (shown in $T$ = 450 fs 2D) is plotted against excitation frequency for different $T$ points (from 100 fs (blue) to 450 fs (red) with 50 fs steps). (c) Area enclosed by the horizontal band and the vertical lines in the $CP_L$ region is integrated and plotted for different $T$. The integrated area is normalized to 1 for the initial $T$ point to directly reflect the $\sim$41\% change. }
 \label{fig3}
\end{figure*}

\subsection*{Coherent 2DES and pump-probe experiments} 
As we noted earlier, the prominent rise and decay dynamics seen in F-2D and F-PP experiments on APC is quite contrasting to that seen in case of LH2. It therefore presents an interesting case using which the assumptions of the $1/N$ limit, namely the infinitely fast EEA compared to any other slower electronic relaxation process, can be better understood. To further investigate this idea, we compare our F-2DES and F-PP experiments to C-2DES and C-PP experiments. The setup details are published elsewhere and described in Figure S3. Figure~\ref{fig3}a shows the real absorptive C-2D spectra of APC at representative pump-probe delays until approximately two times the energy transfer timescale. Compared to the dominant oppositely signed ESA signal in the C-2D spectra of the LH2 protein (see Figure~1 of ref.\cite{Javed2024}, attributed\cite{vanderVegte.2015.J.Phys.Chem.B.119.1302} to the dense two-quantum electronic manifold of collective excitations between strongly coupled sites\sg{)}, only a weak $\sim$9\% ESA feature is seen above $DP_L$. The weak ESA feature is consistent with previous reports\cite{Weng2024,Womick2009,Itakura2025}. Its absence in the F-2D spectra is explained by the mutual cancellation of the ESA contributions\cite{Lott2011,Kunsel2019,Maly2018,Kuhn2020} even if partial (\textit{vide infra}) which further diminishing the net contribution of the inherently weak ESA signals. Identical to the analysis of F-2D spectra in Figure~\ref{fig1}, Figure~\ref{fig3}b shows the band integrated C-2D spectra at a few $T$ slices, and Figure~\ref{fig3}c plots the integrated and normalized $CP_L$ area marked in Figure~\ref{fig3}a. An increase of $\sim$41\% is seen which is quite comparable to the $\sim$44\% increase in the case of F-2D. It should be noted that due to inherently weak ESA signals in APC, the lack of amplitude contrast is also comparable between the C-2D versus F-2D spectra, as compared to the starkly contrasting case of LH2 (see Figure~1 of ref.\cite{Javed2024}). \\

\begin{figure}[h!]
\centering
  \includegraphics[width=7cm]{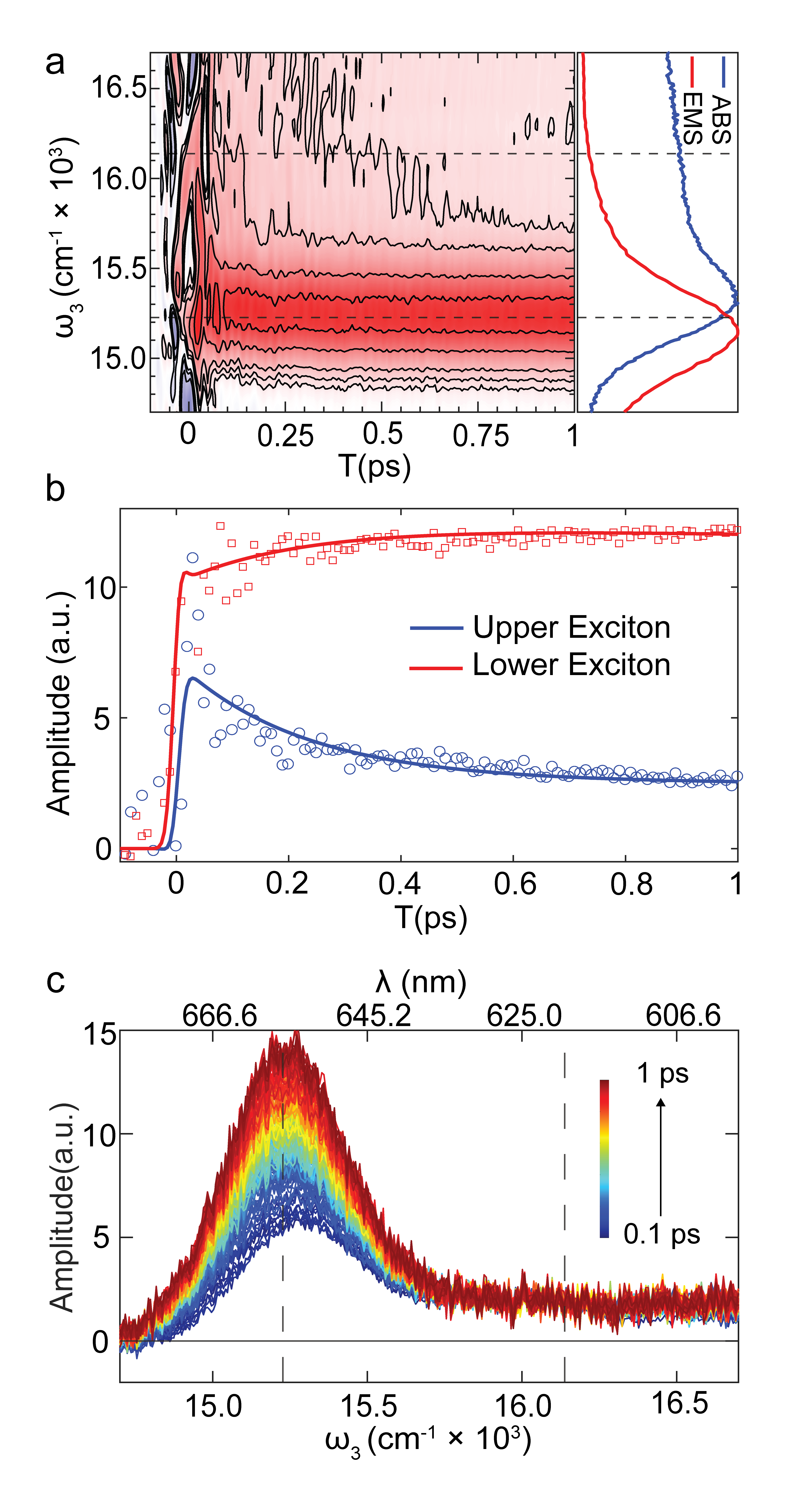}
  \caption{(a) Real absorptive C-PP spectra of APC. Contours are drawn at 10\%, 15\% and from 25 to 100\% in steps of 25\%. Black dashed lines are drawn at 16138 cm$^{-1}$ (620 nm) and 15226 cm$^{-1}$ (656 nm). C-PP spectra peaks at the intersection of ABS (blue) and EMS (red) spectra. (b) The global fit of the C-PP data overlaid at representative wavelength slides showing concomitant rise and decay with a time constant of $224 \pm 3$ fs. (c) C-PP spectra from panel (a) at different $T$ points, normalized at the upper exciton. Blue to red show $T$ range from 0.1 ps to 1 ps in 5 fs step . $\sim$151\% rise observed in the lower exciton region.}
  \label{fig4}
\end{figure}

To extract the kinetic timescales, Figure~\ref{fig4}a shows the C-PP spectra where the maximum passes through the intersection of ABS and EMS spectra. A global kinetic fit with a bi-exponential function yields a $224 \pm 3$ fs time scale for upper to lower exciton energy transfer, in reasonable agreement\cite{Womick2009,Womick2011} with the independently conducted F-PP experiment. Full details of the global analysis are discussed in Section S2. Figure~\ref{fig4}b shows $T$ slices at representative wavelengths along with global kinetic fits to highlight the concomitant rise and decay of the signal, similar to that seen in F-PP spectra (Figure~\ref{fig2}\sg{b}). \\

Figure~\ref{fig4}c shows that the lower exciton region grows by $\sim$151\% compared to $\sim$50\% growth seen in case of F-PP (Figure~\ref{fig2}c). Thus, while the $CP_L$ region in the early $T$ F-2D and C-2D spectra shows comparable growth during energy transfer, when the excitation axis is integrated over in the PP experiment, the overall growth is $\sim$151\% in C-PP versus $\sim$50\% in F-PP. We believe this is caused by two factors – 1.) A larger proportion of the upper exciton is excited in the C-PP experiment. For example, compare 2.7$\times$ larger shoulder height in Figure~\ref{fig2}c versus Figure~\ref{fig4}c. These are overlaid in Figure S4\sg{b}. 2.) The $\sim$9\% negative ESA region near $DP_U$ in C-2D reduces the $DP_U$ amplitude in C-PP spectra thereby amplifyfing the normalized growth (Figure~\ref{fig4}) compared to the F-PP spectra (Figure~\ref{fig2}). 
\FloatBarrier

\subsection*{Coarse-grained F-2D Simulations of APC}

\begin{figure*}[h!]
 \centering
 \includegraphics[height= 13 cm]{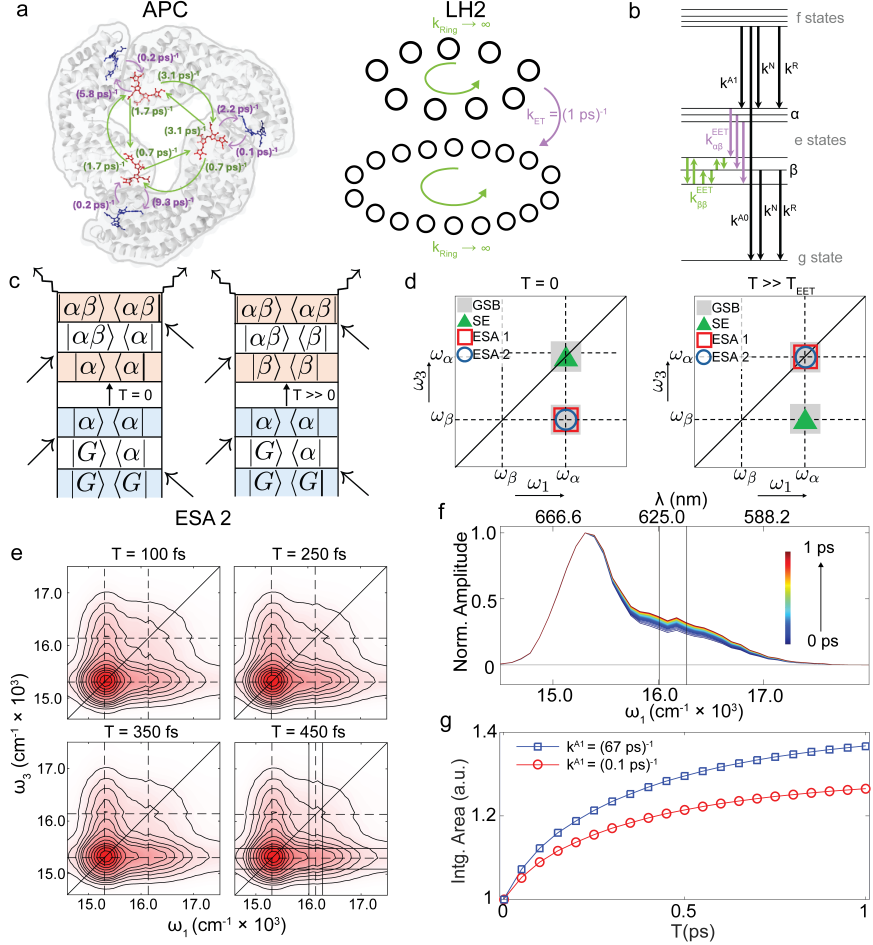}
 \caption{(a) A comparison of electronic equilibration expected in APC (left) versus the LH2 (right) antenna protein. In case of APC, equilibration within $\alpha$ and within $\beta$ sites is significantly slower compared to energy transfer between the nearest neighbor $\alpha-\beta$ sites. The MC-FRET rates\cite{Kunsel2019} (Section S1 and Table S2) account for the relative orientations of the PCB pigments from the APC crystal structure\cite{Kerfeld2022}.  In case of LH2, the $1/N$ limit argument assumes rate of electronic equilibration between within the B800 and B850 manifolds is significantly faster. The B800-B850 energy transfer rates are reported\cite{Novo2003} to have multiple components, all typically < 1 ps.  (b) Rate model for the APC trimer using which the fluorescence quantum yield $Q^{(2)}$ of the ESA2 pathway is estimated (Section S1). $k^{A1}$ denotes the rate of EEA such that two excitations annihilate to leave one excitation in the single-quantum electronic manifold. $k^{A0}$ denotes the event in which no excitation is left behind. $k^R$ and $k^{N}$ denote radiative and additional non-radiative electronic relaxation channels, respectively. $k^{EET}_{ij}$ denotes the rate of EET between sites $i$ and $j$. (c) Double-sided Feynman diagrams for the ESA2 signal pathway in F-2DES which contribute at the $CP_L$ location at $T$ = 0. Each pathway represents a term in the 3$^{rd}$ order perturbative expansion of the time-dependent density matrix element with each light-matter interaction with the laser denoted by arrows. Squiggly lines denote incoherent emission (fluorescence). Time runs vertically such that the space between successive interactions corresponds to time intervals $t_1$, $T$ and $t_3$, respectively. Photon energies corresponding to the first and last interactions determine the position of the pathway on the 2D map in panel (d) along the excitation and detection axis, respectively. With $T \gg 0$, $\alpha \rightarrow \beta$ downhill EET is denoted in the pathways. (d) Positions of the GSB, SE, ESA1, and ESA2 signal pathways in the 2D map for $DP_U$ and $CP_L$ at $T$ = 0 fs and $T \gg T_{EET}$ fs, where $T_{EET}$ denotes the energy transfer timescale. (e) Simulations of F-2D spectra corresponding to the APC model in panels a,b. (f) Analysis of $T$ dynamics, identical to that in Figure~\ref{fig1}, for the simulated F-2D spectra. (g) Relative growth of $CP_{L}$ amplitude of the actual APC model (blue) versus a hypothetical APC model (red) where EEA across all $\alpha$ and $\beta$ sites is significantly faster than other electronic relaxation processes, that is, $k^{A1}$ = (100 fs)$^{-1}$.}
 \label{fig5}
\end{figure*}

Several previous works\cite{Lott2011,Maly2018,Kunsel2019,Kuhn2020} have shown that two types of ESA signals are possible in F-2D - ESA1, which, after 4 interactions, results in a population density matrix element in the one-quantum electronic manifold, such as $\ket{\alpha}\bra{\alpha}$ where site $\alpha$ is excited, and ESA2 which leaves population in a doubly-excited electronic manifold, such as $\ket{\alpha\beta}\bra{\alpha\beta}$ where sites $\alpha$ and $\beta$ are both excited after the action of 4 pulses. The ESA2 pathway is discussed later in Figure~\ref{fig5}c. The argument\cite{Bolzonello2023} for the $1/N$ limit for the ratio of SE/GSB signals essentially relies on the combinatorics of the probability of pump and probe excitations among $N$ isoenergetic sites with perfect electronic mixing such that exciton transfer among the sites, and therefore the EEA rate, is infinitely fast compared to all other radiative or non-radiative electronic relaxation channels. This leads to perfect mutual cancellation of the two ESA signals, which remains so with increasing $N$ or longer pump-probe delays. Extending this argument further, Javed et al.\cite{Javed2024} have shown that strong excitonic couplings and aggregate geometry can render oscillator strength to only a few states, band bottom for J and band top for H aggregates. However, the $1/N$ ratio of SE/GSB signals is only somewhat altered as long as infinitely fast equilibration among the $N$ sites is a valid assumption. \\

Cyanobacterial proteins are an interesting deviation from the above scenario of rapid equilibration within a excitonic manifold. This essential difference is illustrated in Figure~S1 for the case of APC where the PCB pigments associated with the Cys84 amino acid residue in the $\alpha$ and $\beta$ polypeptides are denoted as $\alpha_{84}^i$ and $\beta_{84}^i$, respectively. The superscript $i$ denotes one of the three $\alpha\beta$ protein monomer. In case of APC, the $\alpha_{84}^i-\beta_{84}^j$ bilin chromophores of the adjacent $i$ and $j$ $\alpha\beta$ protein monomers are reported to undergo rapid energy transfer on a timescale\cite{Womick2009,Womick2011,Beck1998} of $\sim$220 fs (Figure ~S1). In comparison, $\alpha_{84}^i-\alpha_{84}^j$, $\alpha_{84}^i-\beta_{84}^i$ and $\beta_{84}^i-\beta_{84}^j$ are only weakly coupled with energy transfer and EEA timescales at low intensities reported to be longer than 67 ps through time-resolved fluorescence anisotropy\cite{Bryant1995} and picosecond transient absorption measurements\cite{ZilinskasDoukas1981,ZilinskasWong1981} on cyanobacterial proteins. Thus, in the APC trimer equilibration within the $\alpha$ or $\beta$ manifold is significantly slower compared to the rapid equilibration\cite{Novo2003} within the B800 and B850 exciton manifolds of the LH2 protein. Additionally, the fluorescence quantum yield of cyanobacterial proteins\cite{Gantt1978} is high – $\sim$0.7 for APC with a fluorescence lifetime\cite{ZilinskasWong1981} of 1.87 ns. This is distinctly different from LH2 where quantum yields of <0.1 \cite{Monshouwer1997} are reported\cite{Pullerits2001}. The above reasoning implies two things - 1. Due to weak coupling between the $\alpha$ and between the $\beta$ sites, the effective $N$ of the $1/N$ limit is reduced approximately to the $\alpha_{84}^i-\beta_{84}^j$ pair, 2.  Radiative electronic relaxation can compete against the slower EEA channels in APC. Note that the above arguments may become stronger when hexameric cyanobacterial proteins such as C-phycocyanin or C-phycoerythrin are considered. \\

\FloatBarrier

Following our previous work\cite{Kunsel2019,Tiwari2018,Migoni2026}, the multi-chromophoric F\"{o}rster resonance energy transfer (MC-FRET) rates in the left panel of Figure~\ref{fig5}a account for the relative orientations of the neighboring PCB pigments as determined by the APC crystal structure\cite{Kerfeld2022}. Further details of the calculations are provided in Section S1. Figure~\ref{fig5}b describes the rate model used to simulate the F-2D spectra. Following the notation from our previous work\cite{Kunsel2019} on the LH2 protein, $k^{A1}$ denotes the rate of EEA such that two excitations in the two-quantum electronic manifold annihilate to leave one excitation in the single-quantum electronic manifold. Similarly, $k^{A0}$ denotes the event in which the two excitations annihilate to leave no excitation behind. This rate is assumed to be zero. $k^R$ and $k^{N}$ denote radiative and additional non-radiative electronic relaxation channels, respectively. $k^{EET}_{ij}$ denotes the rate of energy transfer between sites $i$ and $j$. These processes are denoted in the left panel of Figure~\ref{fig5}a. The key difference\cite{Kuhn2020,Kunsel2019,Maly2018} between the signal pathways which contribute to the C-2D versus F-2D experiments is the last pulse which projects the third-order oscillating non-linear polarization into a population state. For the ESA signals, this leads to multiple pathways resulting from population created on the singly- or the doubly-excited manifold of electronic states and denoted as ESA1 and ESA2, respectively. Each pathway is associated with a corresponding quantum yield, $Q^{(1)}$ and $Q^{(2)}$, respectively, where the latter is sensitive to the rate of EEA. For example, the ESA2 pathway is shown in Figure~\ref{fig5}c. Since the F-2D experiment is blind to the evolution of the resulting two-quantum electronic state after the 4$^{th}$ pulse, the $\alpha$ and $\beta$ excitations can annihilate with rate $k^{A1}$ during this time interval leading to a single-excitation. In one extreme, when EEA rate dominates, $Q^{(2)} = Q^{(1)}$, as was the case for LH2 protein determined in our previous work\cite{Kunsel2019}. In the other extreme when radiative recombination at the individual $\alpha$ and $\beta$ sites dominates,  $Q^{(2)} = 2Q^{(1)}$. Defining response functions as $R_{GSB}$, $R_{SE}$ and $R_{ESA}$ for GSB, SE and ESA signal pathways for the C-2D experiment, the resulting response functions in the F-2D experiment are associated with quantum yields $Q^{(1)}$ and $Q^{(2)}$, and can be written\cite{Kunsel2019} as – 
\begin{align}\label{eqn1}
R_{\mathrm{C2D}} &= R_{\mathrm{GSB}} + R_{\mathrm{SE}} + R_{\mathrm{ESA}}, \nonumber \\
R_{\mathrm{F2D}} &= Q^{(1)} \left( R_{\mathrm{GSB}} + R_{\mathrm{SE}} - R_{\mathrm{ESA}} \right)
                  + Q^{(2)} R_{\mathrm{ESA}} \nonumber \\
                  &= Q^{(1)} \left( R_{\mathrm{GSB}} + R_{\mathrm{SE}}\right) + \left( Q^{(2)} - Q^{(1)}\right) R_{\mathrm{ESA}} \nonumber \\
                 &= Q^{(1)} R_{\mathrm{C2D}} - \left(2 - \frac{Q^{(2)}}{Q^{(1)}}\right) Q^{(1)} R_{\mathrm{ESA}}. 
\end{align}

Note that the above definition is valid in a simpler dimer picture. In general, kinetic rate equations from a model such as in Figure~\ref{fig5}b can be used to derive the expression for $Q^{(2)}$ of the two-quantum electronic state. A detailed derivation has been described for the case of LH2 in our previous work\cite{Kunsel2019, Migoni2026} and modified here for the case of the APC trimer. In the coarse-grained simulations\cite{Migoni2026} of the F-2D spectra based on the response functions described by Eqn.~S3, $Q^{(2)}$ is defined for every pair of molecules fully incorporating the differences in EET rates arising from varying relative orientations across pigment pairs, whereas $Q^{(1)}$ is defined for every molecule. Further details are described in Section S1. \\

Figure~\ref{fig5}c shows that the double-sided Feynman pathways that lead to the $R_{ESA2}$ signal contribution at the $CP_L$ location for a dimer system without any energy shifts in the two electronic quantum manifold. Each such pathway corresponds to the third-order perturbative expansion of the time-dependent density matrix element. Only the rephasing pathway, where the phase of the signal evolution is opposite during $t_1$ and $t_3$, is shown. The photon energies corresponding to the 1$^{st}$ and 4$^{th}$ interaction determines the corresponding spectral location on the 2D map in Figure~\ref{fig5}d. The spectral location is shown for ESA2 as well as the other signal pathways. For the one- and two-quantum electronic manifolds expected in case of the APC protein, the location of SE and ESA signal contributions at $T = 0$ is $DP_U$ and $CP_L$, respectively. With $T$, $\alpha \rightarrow \beta$ energy transfer converts a $\rho_{\alpha\alpha}$ population density matrix element to $\rho_{\beta\beta}$. As shown in Figure~\ref{fig5}d, when energy transfer is complete, that is, $T \gg T_{EET}$, the spectral locations of the SE and ESA signal contributions are swapped. In the context of C-2DES, Figure~\ref{fig5}d suggests that evolving spectral locations of the oppositely signed and comparable $R_{ESA}$ and $R_{SE}$ signal contributions provides enhanced amplitude contrast to the 2D spectra during energy transfer, as observed for the case of the LH2 protein (see Figure~1 of ref.\cite{Javed2024}). However, in case of APC, as noted during the discussion of Figure~\ref{fig3}, the ESA signal is weak to begin with and provides no such contrast during EET even in the C-2D spectra. The dominant ESA signals observed in the C-2D spectra of the LH2 protein originate from strongly coupled sites that give rise to a dense two-quantum electronic manifold of collective transitions with large oscillator strengths. In contrast, the C-2D spectra of the APC trimer exhibit only $\sim$9\% ESA signal amplitude, reflecting comparatively weak intra- and inter-dimer excitonic couplings. \\

In the context of F-2DES, Eqn.~\ref{eqn1} suggests that the two ESA pathways, $R_{ESA1}$ and $R_{ESA2}$ can partially cancel out to reduce the amplitude contrast in the F-2DES spectra. This is even more so when the net ESA signal is dominated by $R_{ESA1}$ pathways which have the same sign as $R_{GSB}$ and $R_{SE}$ pathways, that is, when $Q^{(1)} \ge Q^{(2)}$. For example, Marcus and co-workers\cite{Lott2011,Marcus2012} have reported a 0.31$\times$ reduced strength of $R_{ESA2}$ pathways in (MgTPP)$_2$ dimers due to EEA. Similarly, near perfect cancellation of the two ESA signals in the F-2D spectra \cite{Javed2024,Kunsel2019,Tiwari2018,Karki2019} of LH2 antennas, with $Q^{(2)} \simeq Q^{(1)}$ due to the near perfect yield\cite{Pullerits2001} of EEA, leads to poor contrast\cite{Javed2024} in the F-2D spectra. Considering the APC rate model shown in Figures~\ref{fig5}a,b, the average fluorescence quantum yield $Q^{(2)}$ of the $R_{ESA2}$ pathways is estimated to be $Q^{(2)} \simeq 1.17 Q^{(1)}$ (Table S1) \tlcj{compared to $Q^{(2)} \simeq  1.00 Q^{(1)}$ for the case of the LH2 antenna determined in our previous work\cite{Kunsel2019,Migoni2026}} and 0.31 reported for porphyrin dimers by Marcus and co-workers\cite{Lott2011,Marcus2012}. Figure~\ref{fig5}e presents the coarse grained simulations of the F-2D spectra based on the rate model in Figures~\ref{fig5}a,b. The analysis of the simulated F-2D spectra in Figures~\ref{fig5}f-g is identical to that of the experimental F-2D spectra in Figure~\ref{fig1}. The theoretical F-2D spectra show a growth in the $CP_L$ region that is quite compared to the experiments – $\sim37\%$ and $\sim38\%$, for F-2D and F-PP compared to the experimentally observed growth of $\sim44\%$ and $\sim50\%$, respectively. It should be emphasized that the parameters for the APC rate model in Figures~\ref{fig5}a,b and for estimation of $Q^{(2)} \simeq  1.17 Q^{(1)}$ are directly derived from the reported literature \tlcj{without any adjustable parameters} (Section S1).  A similar analysis in Figure~\ref{fig5}g for a hypothetical case where \tlcj{EEA across all $\alpha$ and $\beta$ sites in the APC trimer is $k^{A1}$ = (100 fs)$^{-1}$, that is, significantly faster than other electronic relaxation processes}, shows estimated $Q^{(2)} \simeq Q^{(1)}$ and $\sim$1.42$\times$ lesser growth in the $CP_L$ amplitude. \\

It should also be noted that the experimentally observed growth in the $CP_L$ region is approximately consistent with a reduced effective system size of $N_{eff}$ $\simeq$ 2 due to negligible inter-dimer coupling in the APC trimer. The resulting $1/N_{eff}$ limit predicts a 50\% relative rise in the $CP_L$ amplitude compared to $\sim$44\% observed in the experiments. The idea of a reduced $N_{eff}$ also explains comparable dynamics between the F-2D and C-2D spectra of the APC trimer versus the striking contrast seen\cite{Javed2024} for the case of the LH2 protein. \\ 

{A modification of the $R_{SE}/R_{GSB} \sim 1/N$ limit combinatorial argument that includes $R_{ESA}$ contributions, by allowing for different strengths of the ESA1 and ESA2 signals through their respective quantum yields $Q^{(1)}$ and $Q^{(2)}$, leads to a $(R_{SE} + R_{ESA})/R_{GSB} \sim \frac{1}{N} \left[ 1 + N (\frac{Q^{(2)}}{Q^{(1)}} - 1) \right]$. The limiting case of perfect EEA is consistent with the $1/N$ limit. For $Q^{(2)} \simeq  1.17 Q^{(1)}$ as theoretically estimated for the case of APC, $(R_{SE} + R_{ESA})/R_{GSB} \sim 0.34$, which is close agreement with the $\sim$37\% growth estimated from the F-2D simulations. Overall, APC represents an interesting deviation from the $1/N$ limit where the total system size can be markedly different from the effective system size that determines early $T$ exciton dymamics in the action-detected spectra.}



\section*{CONCLUSIONS}
In conclusion our results show that the striking contrast between coherent and action-detected 2DES seen for the LH2 protein may be system specific and may not be general for weakly coupled systems. We illustrate this point for the case of APC cyanobacterial protein where comparable growth of the 2D cross peak during energy transfer is seen for coherent and action-detected 2D and pump-probe spectra. Interestingly, compared to the strongly coupled systems such as the LH2 proteins, the ESA spectral band does not represent a dense manifold of collective transitions and therefore the contrast in the 2D peak amplitudes is reduced even for coherent 2D spectra. Simulations of action-detected 2D spectra starting from the APC crystal structure and a zero adjustable parameter rate model confirm these ideas and capture the experiment observed growth in the cross peak amplitude. The combinatorial argument regarding the $1/N$ limit assumes infinitely fast equilibration across sites, and is modified in case of cyanobacterial proteins because of slow equilibration across weakly coupled sites. This reduces the effective system size to $N_{eff} =2$, that is, a dimer, for which the $1/N_{eff}$ argument is still valid. The applied computational scheme can describe more complex situations with subsystems having different quantum yields. Overall, our observations suggest that action detection may be well suited to probe energy transfer across weakly coupled systems.


\section*{ASSOCIATED CONTENT}
\subsection{Supporting Information}
 Description of theoretical method, rate model and simulations of F-2D spectra, experimental setup and analysis details. \\
 

\subsection{Author Contributions}
V.T. and T.L.C.J. conceptualized the project. A.S. built the F-2D setup. A.S. and S.G. performed the initial experiments on the APC trimer. S.G. performed the C-2D experiments on the APC trimer. S.G.M. developed the theoretical framework, performed the F-2D simulations, and wrote the theoretical methods section of the manuscript. V.T. and S.G. wrote the remaining parts of the manuscript with inputs from all authors. 
\subsection{Notes}
The authors declare no competing interests.

\subsection{Data availability}
The data presented in this manuscript are available from the corresponding authors upon reasonable request.
\\
\section*{ACKNOWLEDGEMENTS}

S.G. acknowledges the research fellowship from IISc Bangalore. This work is supported by funding from the Department of Biotechnology, India (No. BT/PR38464/BRB/10/1893/2020). S.G.M. and T.L.C.J. thank the Center for Information Technology of the University of Groningen for their support and for providing access to the Hábrók high-performance computing cluster. This project has received funding from the European Union’s Horizon Europe Research and Innovation Program under the Marie Skłodowska-Curie grant agreement No. 101119442.

\bibliography{references_APC,refsGro} 

\end{document}